# A Hybrid Blockchain-IPFS Solution for Secure and Scalable Data Collection and Storage for Smart Water Meters

Thandile NODODILE[1], Clement NYIRENDA[2]
*Department of Computer Science,
University of the Western Cape,
Bellville, Cape Town, 7535, South Africa*
[1]*Tel: +27 72 711 732, Email: 3692513@myuwc.ac.za*
[2]*Tel: +27 63 930 8916, Email: cnyirenda@uwc.ac.za*

**Abstract:** Scalable and secure data management is important in Internet of Things (IoT) applications such as smart water meters, where traditional blockchain storage can be restrictive due to high data volumes. This paper investigates a hybrid blockchain and InterPlanetary File System (IPFS) approach designed to optimise storage efficiency, enhance throughput, and reduce block time by offloading large data off-chain to IPFS while preserving on-chain integrity. A substrate-based private blockchain was developed to store smart water meter (SWM) data, and controlled experiments were conducted to evaluate blockchain performance with and without IPFS. Key metrics, including block size, block time, and transaction throughput, were analysed across varying data volumes and node counts. Results show that integrating IPFS significantly reduces on-chain storage demands, leading to smaller block sizes, increased throughput, and improved block times compared to blockchain-only storage. These findings highlight the potential of hybrid blockchain-IPFS models for efficiently and securely managing high-volume IoT data.

## 1. Introduction

In the digital era, blockchain technology has shown immense potential to revolutionise data management by providing decentralised, secure and immutable storage. This concept gained widespread attention following the publication of the foundational Bitcoin whitepaper by Nakamoto [1]. Blockchain combines various software and theoretical mechanisms, including cryptography, networking, hashing algorithms, and a unique storage model, to achieve decentralisation, peer-to-peer interaction, immutable storage and enhanced security. Data in the blockchain is organised into blocks that are sequentially created, appended, and cryptographically linked through hashes generated from the root hash of the previous block [2].

Research has shown promising applications for blockchain in data-intensive domains. For instance, Wenjun et al. [3] applied blockchain technology to optimise intelligent water management. However, as data-intensive applications like IoT networks, environmental monitoring, and financial transactions continue to grow, blockchain's limitations, particularly in storage capacity and latency, have become apparent [4], [5], [6]. For example, studies on blockchain-based smart water meter systems [7] highlight that constraints on block storage size can lead to bottlenecks and latency during block creation and finalisation. A promising approach to address these storage limitations is through off-chain storage solutions. Yang et al. [2] analysed different peer-to-peer (P2P) data networks such as Nasper [8], BitTorrent [9], Swarm [10], Storj [11] and the Hypercore

Protocol [12] and noted that IPFS is most utilised in the blockchain. P2P networks provide scalable storage alternatives, and among these, the InterPlanetary File System (IPFS) [13] has emerged as a popular standard and optimised choice for blockchain integration [2]. IPFS is a peer-to-peer distributed file system that optimises data storage by using content-addressable data hashes. This system allows blockchain applications to manage high-frequency data more efficiently by offloading bulk storage requirements. Despite IPFS's potential to address blockchain's scalability and storage challenges, there is a limited quantitative assessment of its impact on key performance metrics like block time, block size, and throughput (measured as transactions per second).

This study aims to address this research gap by evaluating the effects of IPFS-enabled blockchain performance under fixed node and fixed data conditions. The study examines whether the IPFS solution significantly influences these critical metrics, which are essential for applications that require high throughput and prompt data confirmation. To conduct this analysis, a private blockchain network was developed using the Substrate framework, with sampled digital water meter data stored on the blockchain via extrinsic call events.

The rest of the paper is organised as follows: Section 2 outlines the research objectives, while Section 3 describes the methodology used in the experiments. Section 4 discusses the technology and business implications of the findings. Section 5 presents the results and analysis, followed by section 6, which explores the business benefits of the hybrid blockchain-IPFS solution for secure and scalable data collection and storage in smart water meters. Finally, section 7 concludes the paper.

## 2. Objectives

The main objectives of this research are to:
- Measure and compare Substrate-based blockchain performance metrics (block time, block size and transaction per second as throughput) with and without IPFS under controlled conditions.
- Investigate and assess the scalability limits of on-chain-only storage and demonstrate how IPFS integration improves performance.
- Assess the impact of IPFS on reducing latency and bottlenecks in the blockchain-based smart water meter system.
- Determine the significance of performance differences using t-tests.

## 3. Methodology

*3.1 – Overview of Experimental Design*

The evaluation of this work consists of a private substrate-based blockchain network with ten substrate customised nodes built by Parity and open-source community [14], nine as validator nodes and one node for genesis configuration. The validator nodes produce and finalise blockchain blocks. The blockchain nodes have been modified to add the functionality of storing data on-chain using a pallet. IPFS desktop node application has been installed from the official website [15]. The required specifications for executing private blockchain are computer memory of at l6 GB RAM, available storage of at least 10 GB, and broadband internet connection for the MacOS operating system with Apple silicon required the software development components: Protobuf, OpenSSL, Cmake, and Rust. The smart water meter data had been sourced from the Queensland Government open data portal, which is generated from residential homes and commercial properties [16]. The experiment was carried out over 3 days with 90 data captures after every block finalised and storing of on-chain data hashes.

*3.2 – Test data*

The Unitywater Digital Water Meter data is 1.27 GB in size and collected over 3 months, 90 data files each with approximately 230,000 records in each file. To assess for large data storage, the same dataset was sampled to achieve 3960 data files of 56.38 GB in total.

*3.3 – Experimental Runs With and Without IPFS*
   a) With IPFS:
      - Fixed Node Count and Variable Data Size: Execution of 5 simulations at ten nodes, increasing data hashes from 10 to 90 to assess how IPFS handles growing off-chain data.
      - Fixed Data Size and Variable Node Count: Fixed data hashes at 90 and varying nodes from 3 to 10, examining how IPFS scales with more nodes.
   b) Without IPFS:
      - Fixed Node Count and Larger Data Load: 10 nodes and scaled data hashes significantly from 50 up to 800 to simulate large raw data on-chain,
      - Fixed Large Data Load and Variable Node Count: Fixed data load at 800 hashes varied nodes from 3 to 10.

Applying Python library SciPy [17] python library for scientific computation, t-tests were performed for non-IPFS private blockchain, with a fixed node count of ten and fixed data hashes at 800. And with IPFS-enabled private blockchain with a fixed node count of 10 and fixed data hashes at 800 bytes. The aim is to obtain t-statistics p-values to understand the statistical significance of the data output of the metrics. The significance of experimental options for the fixed nodes and increasing data presents practical IoT data growth in the blockchain where the number of nodes remains relatively stable. With IPFS, it tests whether the blockchain can maintain consistent performance with rising data volumes. For the fixed data and increasing nodes, the setup tested blockchain scalability, assessing if additional nodes help to offset the storage burden. The non-IPFS simulations demonstrate whether more nodes lead to diminishing returns with the performance metrics or not.

## 4. Technology or Business Case

This study explores the technological and business implications of integrating InterPlanetary File Systems (IPFS) with blockchain, specifically targeting use cases that require efficient data storage, retrieval, and transaction throughput. IPFS provides a distributed, content-addressed storage model that, when combined with blockchain, enhances scalability by reducing the amount of data directly stored on-chain [2]. The technological benefits of this hybrid integration are significant for industries where data integrity, security, and efficiency are crucial. IPFS's functionality to store data off-chain while providing unique data hashes enables blockchain networks to manage large data volumes without impacting network performance [18]. This work focuses on the analysis of IPFS-enabled and non-IPFS blockchain setups to assess improvements in block size, block time, and throughput.

However, despite its advantages, the integration of IPFS in distribution networks presents several challenges that need to be addressed for practical deployment.
- Data Persistence and Availability: IPFS operates on a peer-to-peer basis, meaning that files are not stored permanently unless they are pinned by nodes. Without proper data persistence strategies, critical data may become unavailable if it is not actively hosted by nodes, requiring additional storage incentives or off-chain backup solutions to ensure long-term accessibility.

- Scalability and Computational Overhead: Although IPFS offloads large data from the blockchain, ensuring efficient retrieval at scale requires adequate infrastructure support. Large-scale adoption of hybrid blockchain blockchain-IPFS solutions may necessitate optimisations such as caching mechanisms, decentralised content delivery strategies, or integration with emerging IPFS enhancements like Filecoin-based incentivization [19].

The hybrid solution remains highly advantageous in sectors such as finance, supply chain, and healthcare, where high-frequency transactions and verifiable data integrity are crucial. It reduces on-chain storage costs, improves scalability, and enhances accessibility for data-intensive applications.

## 5. Results And Analysis

This section presents and analyses the experimental outcomes, comparing the private substrate blockchain performance with and without IPFS integration considering the metrics block time, block size, and throughput as transaction per second (TPS) as established in the methodology.

*5.1 – Experimental Runs With and Without IPFS*

To provide a comprehensive view, we conducted simulations with both IPFS-enabled and non-IPFS setups. The experimental results, visualised in figures 1,2,3 and 4, illustrate the effects of IPFS on the key metrics of block time, block size, and throughput considering the two scenarios: fixed data hashes and fixed number of nodes. The graphs in Fig.1 show the metrics of block time and data hashes, block size and data hashes, and throughput and data hashes with a fixed number of nodes at 10 for IPFS-enabled private blockchain based on Substrate.

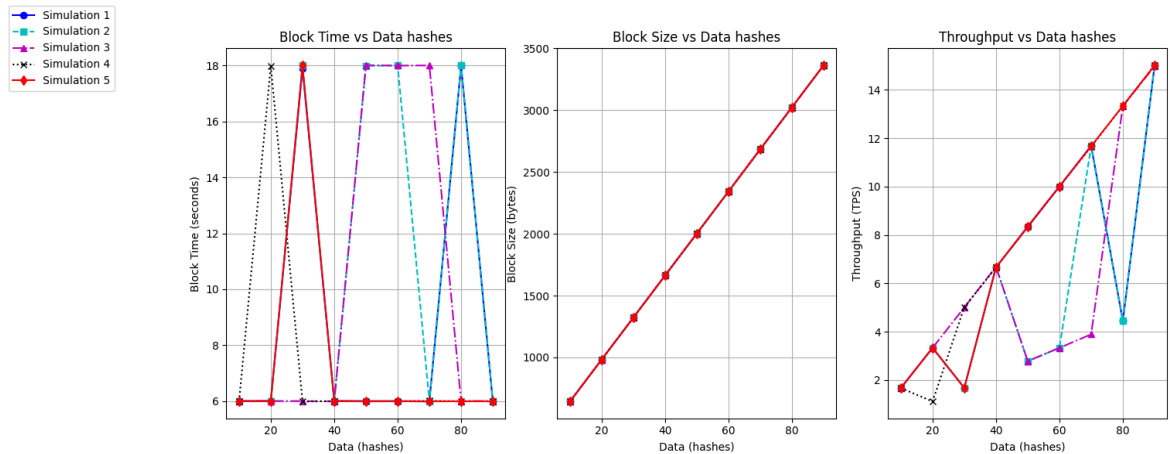

*Figure 1: IPFS-Enabled Simulation Results – Block Time, Block Size, and Throughput vs Data Hashes*

The graph of block time vs data hashes shows that generally, block time increases with more data hashes, but the relationship is not always linear. The different simulations show varying block time with a minimum of 6 seconds and 18 seconds.
The graph of block size versus data hashes shows the relationship between the size of each block (in bytes) and the number of data hashes included. Block size grows proportionally with the number of data hashes. All simulations 1,2,3,4, and 5 show a consistent trend in the upward trajectory. The graph of throughput (transaction per second) versus data hashes per block is finalised. The TPS increases with more data hashes; this indicates improved system efficiency. All simulations show distinct throughput curves, however, in the upward trajectory. The graphs in Fig.2 show the metrics of block time and data hashes, block size and data hashes, and throughput and data hashes with a fixed number of nodes at 10 for non-IPFS private blockchain.

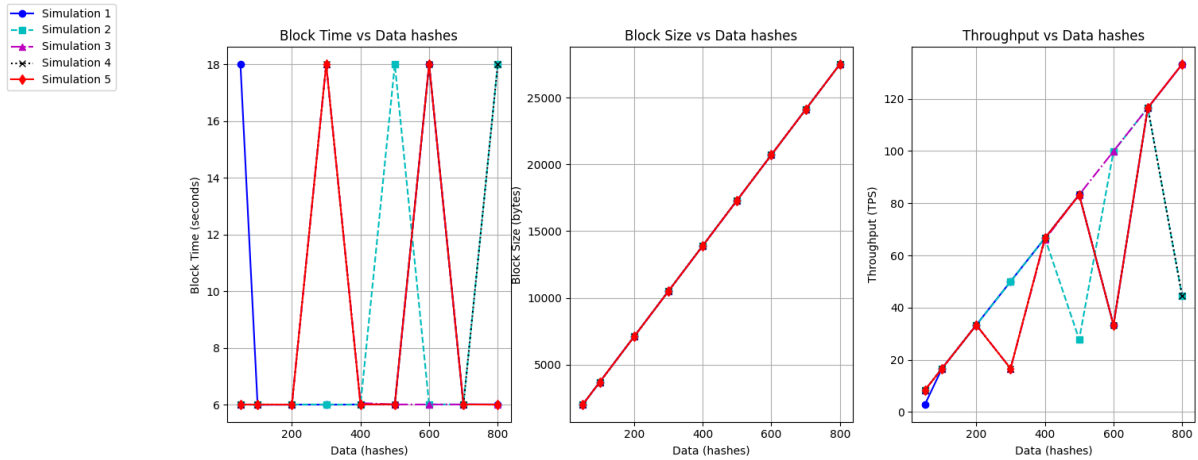

*Figure 2: Non-IPFS Simulation Results – Block Time, Block Size, and Throughput vs Data Hashes*

The resultant outputs resemble Fig.1; however, data hashes have been drastically increased from 50,100…,800 to simulate large data. The block size gradually increases as more data hashes are added on-chain. The throughput shows that it is not necessarily affected by the increase in data size as there is an increase as more data hashes increase. With regards to block time and data hashes, all different simulations show varying block times. The graphs in Fig.3 show the metrics of block time and number of nodes, throughput and number of nodes with fixed data hashes at 90 for IPFS-enabled private blockchain.

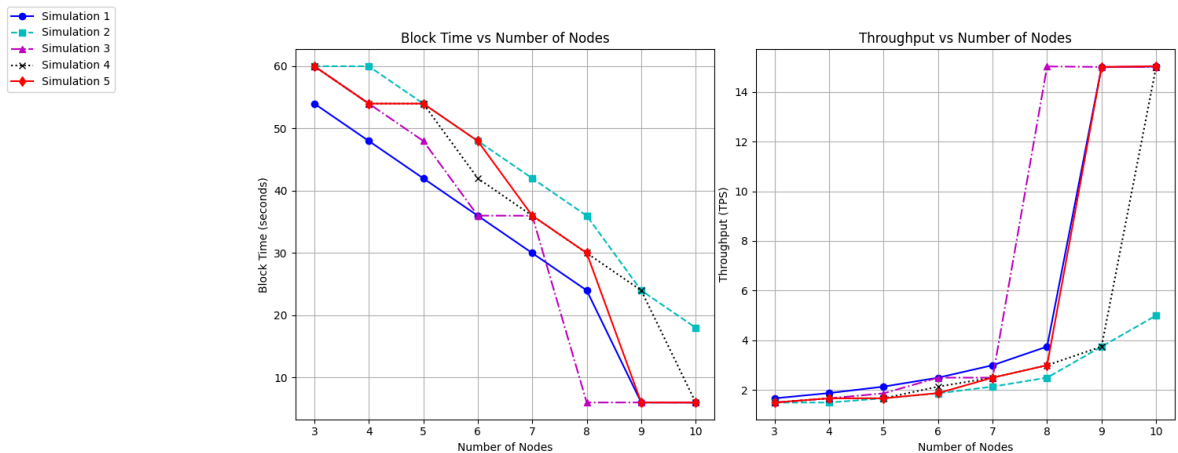

*Figure 3: IPFS-Enabled Simulation Results – Block Time, and Throughput vs Number of Nodes*

The time it takes to process transactions in a block decreases significantly as more nodes are added to the private blockchain for all simulations. This is prominent due to the reduction of block creation and finalisation by validation nodes. In the graph of throughput and node counts, as more nodes are added to peer and connect to the blockchain, the throughput increases even with large data size.

The graphs in Fig.4 show the metrics of block time and number of nodes, throughput, and number of nodes with a fixed number of data hashes at 800 for non-IPFS private blockchain. The block time decreases as more nodes are added to the blockchain. However, there is a slow, gradual increase in throughput due to large data not being stored on IPFS. This shows that the blockchain takes much time to process the blocks even when more nodes are being added.

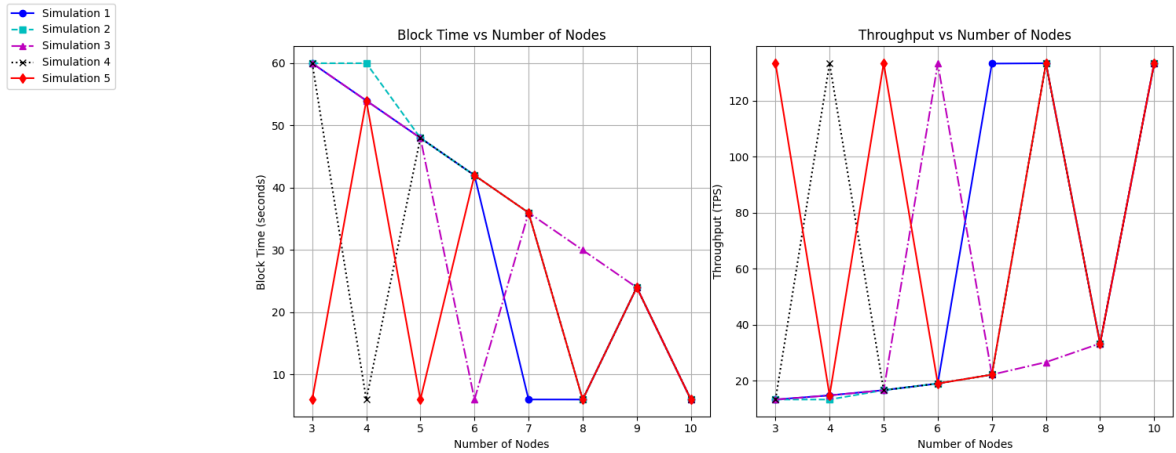

*Figure 4: Non-IPFS Simulation Results – Block Time and Throughput vs Number of Nodes*

The findings show that with IPFS as offline storage, reducing the storage capacity of the data stored on-chain significantly affects the block time to decrease and throughput to increase for more capacity during block creation and finalisation.

*5.2 – T-tests on Block Time, Block Size, and Throughput*

To determine the statistical significance of IPFS's impact between the IPFS-enabled and non-IPFS configurations, we performed independent t-tests for the metrics of block time, block size, and throughput. The results are outlined below after application of the t-test formula above:

*5.2.1 – Fixed Nodes Analysis*

- Block Time: The t-test yielded a t-statistic of 0.2414 with a p-value of 0.8098, indicating no statistically significance difference in block time between the configurations with and without IPFS.
  The results indicate that, under fixed nodes, whether with or not IPFS, there is no impact on the time taken to add or create a block.
- Block Size: A significant difference was observed in the block size after the application of the t-test formula, with a t-statistic of -9.3864 and a p-value less than 0.0001. The data shows that IPFS usage correlates with a reduction in block size compared to a configuration without IPFS. This reduction results from IPFS's efficient content-addressing.
- Throughput: The t-test for throughput yielded a t-statistic of -8.0768 and a p-value below 0.0001, indicating a statistically significant difference between the IPFS and non-IPFS configuration. This shows that incorporating IPFS is associated with high throughput. IPFS allows more transaction operations within each block interval after block size reduction. This result suggests that IPFS integration or solution can enhance network efficiency, even when node count remains fixed.

- *5.2.2 – Fixed Data Hashes Analysis:*
  - Block Time: Under fixed data hashes yielded a t-statistic of 1.3592 with a p-value of 0.1780, indicating no statistically significance difference between with and without IPFS. This implies that the IPFS integration does not affect block validation time when data hashes remain fixed.

The findings highlight that IPFS integration offers substantial advantages in terms of scalability and throughput without compromising block time. Thus, the hybrid blockchain-IPFS systems are more likely to be effective for data-intensive applications where network efficiency, storage optimisation, and scalability are paramount. This analysis serves as a foundation for future implementations of blockchain and IPFS in distributed, high-throughput applications and underscores the practical advantages of combining blockchain with IPFS for efficient data management.

## 6. Business Benefits

The integration of IPFS with blockchain offers several business advantages that improve the efficiency, scalability and cost-effectiveness of data management:
- Reduced Storage Costs: By leveraging IPFS's content-addressing optimisation mechanism, businesses can minimise redundant data storage on-chain, which decreases overall storage costs.
- Increased Transactional Throughput: With IPFS reducing block size, blockchain networks can support a high volume of transactions per block. Thus, this enables businesses and individuals to scale operations efficiently, making blockchain solutions more viable for high-frequency transactions.
- Improved Blockchain System Performance: The offloading of data storage onto IPFS reduces network congestion and significantly improves the overall performance of blockchain systems, allowing faster and more reliable operations.

While hybrid blockchain-IPFS models present significant benefits, real-world deployment faces several challenges:
- Network Latency and Data Retrieval Times: IPFS retrieval speed depends on node availability and network connectivity, which introduces unpredictable delays in accessing stored data [21].
- Security and Access Control: Ensuring access to IFPS-stored data requires additional cryptographic mechanisms and access control policies to prevent unauthorised modifications [22].
- Compliance with Regulations: Industries handling sensitive data must ensure compliance with legal and data protection regulations, which may delay the implementation and deployment.

A phase implementation strategy is recommended, with short short-term proof-of-concept (POC) trials between 1 to 5 weeks, followed by a pilot study of about 2 months before full-scale deployment 2 to 3 months, depending on industry requirements and infrastructure readiness.

## 7. Conclusions

This paper explores a hybrid blockchain-IPFS solution for secure and scalable data collection and storage in smart water meters. It evaluates private substrate blockchain configurations with and without IPFS integration, focusing on performance metrics: block time, block size, and throughput. The results demonstrate that while IPFS integration does not significantly impact block time, it effectively reduces block size and enhances throughput. This demonstrates that IPFS not only reduces storage burdens but also accelerates processing, making it an effective solution for real-world applications that demand both security and scalability.

In summary, the IPFS and blockchain integration presents a powerful approach that redefines data management within the blockchain, making it an invaluable tool for both developers and businesses aiming for high-performance and future-proof solutions. However, this approach has limitations, including the potential latency in retrieving IFPS-stored data due to network dependencies, and the need for additional mechanisms to ensure long-term data availability. Despite these challenges, this study offers a foundation for future work and refinement of the IPFS-based model to expand blockchain's reach and usability in data-heavy sectors.

## Declaration of use of content generated by Artificial Intelligence (AI) (including but not limited to Generative-AI) in the paper

The authors acknowledge the use of content generated by Artificial Intelligence (AI) (including but not limited to text, figures, images, and code) in the paper entitled "**A Hybrid Blockchain-IPFS Solution for Secure and Scalable Data Collection and Storage in Smart Water Meter**" using Grammarly and ChatGPT to improve the grammar of the paper.